\documentclass[12pt]{article}
\pdfoutput=1
\usepackage{eucal}

\DeclareFontFamily{T1}{calligra}{}
\DeclareFontShape{T1}{calligra}{m}{n}{<->s*[1.44]callig15}{}
\DeclareMathAlphabet\mathcalligra   {T1}{calligra} {m} {n}
\DeclareMathAlphabet\mathzapf       {T1}{pzc} {mb} {it}
\DeclareMathAlphabet\mathchorus     {T1}{qzc} {m} {n}
\DeclareMathAlphabet\mathrsfso      {U}{rsfso}{m}{n}
\DeclareMathAlphabet\mathfrcal      {T1}{frcursive}{m}{it}
\DeclareFontFamily{T1}{frcursive}{}
\DeclareFontShape{T1}{frcursive}{m}{n}{<->s*[1.44]callig15}{}
\DeclareMathAlphabet\mathfrcal      {T1}{frcursive}{m}{it}

\usepackage{amsmath}
\usepackage{amssymb}
\usepackage{graphicx}
\usepackage[dvipsnames]{xcolor}
\numberwithin{equation}{section}
\usepackage{array}
\usepackage{mathtools}
\usepackage{dsfont}
\usepackage{mathrsfs}
\usepackage{tikz}\usetikzlibrary{matrix,fit}
\usepackage{varwidth}
\usepackage{enumerate}

\usepackage[margin=1in]{geometry}

\usepackage{mathabx}
\usepackage{empheq}

\usepackage{hyperref}
\hypersetup{
    colorlinks=false,
    linkbordercolor	=BlueViolet,
citebordercolor	=OliveGreen,
urlbordercolor=RoyalBlue
}
\usepackage{setspace}

\usepackage[
    backend=bibtex,
    style=alphabetic,
maxbibnames=99,
giveninits=true,
minalphanames=1,
maxalphanames=3, doi=false,isbn=false,url=false
  ]{biblatex}

\bibliography{PolivanovBibShort}

\renewbibmacro{in:}{}

\newbibmacro{string+doi}[1]{%
  \iffieldundef{doi}{#1}{\href{http://dx.doi.org/\thefield{doi}}{#1}}}
\DeclareFieldFormat{title}{\usebibmacro{string+doi}{\mkbibemph{#1}}}
\DeclareFieldFormat[article]{title}{\usebibmacro{string+doi}{\mkbibquote{#1}}}

\usepackage{slashed}
\usepackage{upgreek}
\usepackage{appendix}

\usepackage{tabstackengine}
\fixTABwidth{T}

\newdimen\mytextwidth
\newcommand\rem[2][cyan!40!green]{\noindent\nobreak\hfil\penalty1000\hfilneg
\mytextwidth=\linewidth\advance\mytextwidth by 2mm%
\begin{tikzpicture}[baseline=-\the\dimexpr\fontdimen22\textfont2\relax]\node[outer sep=0pt,draw=black,fill=#1,fill opacity=1,text opacity=1,rectangle,rounded corners]{\begin{varwidth}{\mytextwidth}\textcolor{white}{#2}\end{varwidth}};
\end{tikzpicture}\allowbreak%
}

\newcommand\whiterem[2][white!]{\noindent\nobreak\hfil\penalty1000\hfilneg
\mytextwidth=\linewidth\advance\mytextwidth by 2mm%
\begin{tikzpicture}[baseline=-\the\dimexpr\fontdimen22\textfont2\relax]\node[outer sep=0pt,draw=black,fill=#1,fill opacity=1,text opacity=1,rectangle,rounded corners,line width=1.5pt]{\begin{varwidth}{\mytextwidth}\textcolor{black}{#2}\end{varwidth}};
\end{tikzpicture}\allowbreak%
}

\newcommand{\dd}{\partial}
\newcommand{\bd}{\overline{\partial}}
\newcommand{\CP}{\mathds{CP}}
\newcommand{\CC}{\mathds{C}}

\renewcommand{\bar}{\overline}
\renewcommand{\tilde}{\widetilde}

\newcommand{\bea}{\begin{equation}}
\newcommand{\eea}{\end{equation}}
\newcommand{\bear}{\begin{eqnarray}}
\newcommand{\eear}{\end{eqnarray}}
\newcommand{\bearr}{\begin{eqnarray*}}
\newcommand{\eearr}{\end{eqnarray*}}

\newcommand{\appendixnumberline}[1]{Appendix #1.\space}

\let\oldappendix\appendix
\makeatletter
\renewcommand{\appendix}{%
  \addtocontents{toc}{\let\protect\numberline\protect\appendixnumberline}%
  \renewcommand{\@seccntformat}[1]{\large\bfseries Appendix . }%
  \oldappendix
}
\makeatother

\usepackage{mdframed}

\newmdenv[
  topline=false,
  bottomline=false,
  rightline=false,
  linewidth=2pt,
  skipabove=\topsep,
  skipbelow=\topsep
]{siderules}

\newmdenv[
  topline=false,
  bottomline=false,
  linewidth=2pt,
  skipabove=\topsep,
  skipbelow=\topsep
]{siderulesright}

\makeatletter
\renewcommand{\@seccntformat}[1]{\csname the#1\endcsname.\quad}
\makeatother

\newcommand\scalemath[2]{\scalebox{#1}{\mbox{\ensuremath{\displaystyle #2}}}}

\usepackage{setspace}
\usepackage{xpatch}

\makeatletter
\renewcommand{\@chap@pppage}{%
  \clear@ppage
  \thispagestyle{plain}%
  \if@twocolumn\onecolumn\@tempswatrue\else\@tempswafalse\fi
  \null\vfil
  \markboth{}{}%
  {\centering
   \interlinepenalty \@M
   \normalfont
   \MakeUppercase \appendixpagename\par}%
  \if@dotoc@pp
    \addappheadtotoc
  \fi
  \vfil\newpage
  \if@twoside
    \if@openright
      \null
      \thispagestyle{empty}%
      \newpage
    \fi
  \fi
  \if@tempswa
    \twocolumn
  \fi
}
\makeatother

\usepackage{tocloft}

\let \savenumberline \numberline
\def \numberline#1{\savenumberline{#1.}}

\usepackage{etoolbox}
\patchcmd{\tableofcontents}{\@starttoc}{\vspace{-0.3cm}\@starttoc}{}{}

\usepackage{hyperref}

\usepackage{titlesec}

\titleformat*{\section}{\large\bfseries}
\titleformat*{\subsection}{\normalsize\bfseries}
\titleformat*{\subsubsection}{\normalsize\bfseries}
\titleformat*{\paragraph}{\large\bfseries}
\titleformat*{\subparagraph}{\large\bfseries}
\titlespacing{\author}{-5pt}{-5pt}{-5pt}[-5pt]

\makeatletter
\renewcommand\subsubsection{\@startsection{subsubsection}{3}{\z@}%
                                     {-3.25ex\@plus -1ex \@minus -.2ex}%
                                     {-1.5ex \@plus -.2ex}
                                     {\normalfont\normalsize\bfseries}}
\renewcommand\subsection{\@startsection{subsection}{3}{\z@}%
                                     {-3.25ex\@plus -1ex \@minus -.2ex}%
                                     {-1.5ex \@plus -.2ex}
                                     {\normalfont\normalsize\bfseries}}                                     
\makeatother

   \usepackage{stmaryrd}

    \makeatletter
\def\@fnsymbol#1{\ensuremath{\ifcase#1\or \ast \or \ast\ast\or\ast \or \dagger\or \ddagger\or
   \mathsection\or \mathparagraph\or \|\or **\or \dagger\dagger
   \or \ddagger\ddagger \else\@ctrerr\fi}}
    \makeatother
    
    \newcommand{\tens}[1]{%
  \mathbin{\mathop{\otimes}\limits_{^{\textrm{\normalsize #1}}}}%
}

\begin{document}

\title{\vspace{-1.0cm} Sigma models as Gross-Neveu models\footnote{Contribution to the proceedings of the conference ``Polivanov-90'' in memory of M.~K.~Polivanov, held on 16-17 December 2020 at the Steklov Mathematical Institute in Moscow.}}
\author{Dmitri Bykov\thanks{Emails:
bykov@mi-ras.ru, dmitri.v.bykov@gmail.com}
\\  \vspace{-0.3cm}  \\
{\small Steklov Mathematical Institute of Russian Academy of Sciences}\\\vspace{-0.9cm} \\ {\small  8 Gubkina St., Moscow 119991, Russia \;}
}

\date{}

{\let\newpage\relax\maketitle}

\maketitle

\vspace{-0.5cm}
\textbf{Abstract.}
We review the correspondence between integrable sigma models with complex homogeneous target spaces and chiral bosonic (and possibly mixed bosonic/fermionic) Gross-Neveu models. Mathematically, the latter are models with quiver variety phase spaces, which reduce to the more conventional sigma models in special cases.  We discuss the geometry of the models, as well as their trigonometric  and elliptic deformations, Ricci flow and the inclusion of fermions.

\vspace{1cm}

In a series of recent papers~\cite{BykovNilp, BykovGN, BykovSUSY} (see also the review~\cite{BykovAffleck}) we proposed an equivalence between integrable sigma models with complex homogeneous target spaces and chiral bosonic (and mixed bosonic/fermionic) Gross-Neveu models. Typical target spaces of models in this class include the projective space~$\CP^{n-1}$, Grassmannians and flag manifolds, but the formulation in terms of the Gross-Neveu model suggests that it is more appropriate to speak of the phase space as a fundamental object rather than the target (configuration) space. The admissible phase spaces are conjecturally quiver varieties (see~\cite{Nakajima} for the definition) and are more general than cotangent bundles such as $T^\ast \CP^{n-1}$. 

Apart from broadening the class of integrable models, this reformulation sheds new light on various long-standing questions related to sigma models. First and foremost, it turns out that all non-linear constraints typical for sigma models can be effectively solved in the new formulation, and, surprisingly, the interactions in the models are polynomial. The $\beta$-function of the theory is described by elementary diagrams typical of Gross-Neveu models (in the deformed case as well!), and the computation does not require the background field method. Moreover, the universality of the $\beta$-function for all symmetric spaces with a given symmetry group can be easily explained and extrapolated to non-symmetric spaces as well. We have also conjectured that the so-called Yangian anomalies (anomalies in L\"uscher's non-local charges, see the book~\cite{AbdallaBook} for a review) are in fact related to the much more familiar chiral anomalies in the Gross-Neveu model. The proposed formalism allows to incorporate fermions in a natural way as well, both supersymmetrically and non-supersymmetrically. In the former case the approach based on Gross-Neveu models provides a formulation of models with worldsheet supersymmetry purely in terms of target space super-geometry, using the concept of supersymplectic quotient.

In the present note we review and extend some of these results. In section~\ref{CPnGNsec} we start with the $\CP^{n-1}$ model, where the relation to bosonic Gross-Neveu models can be motivated and explained most easily. As an elementary application, in section~\ref{QMsec} we consider the quantum-mechanical reduction of the $\CP^{n-1}$ model showing that the Gross-Neveu formalism provides a manifest $SU(n)$-invariant expression for the Laplace-Beltrami operator. We then explain in section~\ref{compsympsec}, why complex symplectic geometry naturally enters the game, and how this point of view momentarily hints at a wide class of models amenable to a similar analysis.  Finally, in section~\ref{defsec} we demonstrate that the integrable deformations may be obtained by inserting the classical $r$-matrix in the quartic interaction of the Gross-Neveu model, and we conclude in section~\ref{fermionsec} with the discussion of the inclusion of fermions in models of this type.

\section{The $\CP^{n-1}$-model as a gauged chiral Gross-Neveu model}\label{CPnGNsec}

To motivate the relation between sigma models and Gross-Neveu models it is easiest to consider a concrete  example -- the $\CP^{n-1}$-model, which exhibits all the salient features of the correspondence. We start with the $n$-flavor chiral Gross-Neveu model with coupling constant $\kappa$ ($\gamma_5=\sigma_3$ is the notation we use in 2D):
\bea\label{lagr1}
\mathscr{L}=\sum\limits_{a=1}^n\,\bar{\Psi_a} \slashed{D} \Psi_a + \kappa\,\sum\limits_{a, b=1}^n\,\left(\bar{\Psi_a}{1+\gamma_5\over 2}\Psi_a\right) \,\left(\bar{\Psi_b}{1-\gamma_5\over 2}\Psi_b\right)
\eea
The Dirac `spinor' $\Psi$ may be decomposed in its Weyl components in the standard way:
\bea\label{psidecomp}
\Psi_a=\begin{pmatrix} U_a\\ \bar{V}_a\end{pmatrix}\,,\quad\quad a=1, \ldots, n
\eea
The crucial point is that we will assume $\Psi_a$'s to be \emph{bosonic} variables, which is a clear distinction from the fermionic Gross-Neveu models studied in the 70's~\cite{GrossNeveu, WittenThirring}  (mixed bosonic/fermionic systems are discussed below in section~\ref{fermionsec}). To see the relation between~(\ref{lagr1}) and the sigma model, we rewrite the Lagrangian in terms of the $U$- and $V$- variables:
\bea\label{lagr2}
\mathscr{L}=\sum\limits_{a=1}^n\,\left(V_a \cdot \bar{D} U_a+\bar{U}_a \cdot D \bar{V}_a\right) +\kappa\,\left(\sum\limits_{a=1}^n\,|U_a|^2\right) \left( \sum\limits_{b=1}^n\,|V_b|^2\right)\,.
\eea
`Chirality' of the model has to do with its invariance w.r.t. the complex transformations
\bea
U_a\to \uplambda\,U_a,\quad\quad V_a\to \uplambda^{-1}\,V_a,\quad\quad\textrm{where}\quad\quad \uplambda \in \CC^\ast\,.
\eea
It can be traced back to the presence of the chiral projectors ${1\pm \gamma_5\over 2}$ in the Lagrangian~(\ref{lagr1}). The subgroup $U(1)\subset \CC^\ast$ corresponds to the `vectorial' transformations $\Psi_a \to e^{i\,\upalpha} \Psi_a$. This is also a good point to introduce the covariant derivatives entering the above Lagrangians:
$
\bar{D} U_a={\dd \over \dd \bar{z}} U_a-i\,\bar{\mathcal{A}}\,U_a
$. Formally speaking, these are~$U(1)$ covariant derivates, however they ensure classical invariance w.r.t. the enlarged~$\CC^\ast$ symmetry, if one assumes the transformation property $\bar{\mathcal{A}}\to \bar{\mathcal{A}}-i\, \bd \log{\uplambda}$. As this is a chiral symmetry, typically it will be anomalous at the quantum level~\cite{BykovGN}.

An important fact about the Lagrangian~(\ref{lagr2}) is that it is quadratic in the $V$-variables (as well as in the $U$-variables, since there is the apparent symmetry ${U\leftrightarrow \bar{V}}$, ${z\leftrightarrow \bar{z}}$), so that we can eliminate them via the e.o.m. As a result, we obtain the Lagrangian
\bea\label{lagr3}
\mathscr{L}={1\over \kappa}\,\frac{|\bar{D} U|^2}{|U|^2}\,.
\eea
This is nothing but a very concise form of the $\CP^{n-1}$-model. To see the relation to the more conventional formulations, recall the local projective invariance $U_a\to \uplambda\,U_a$. Using it, we may choose a partial gauge $\sum\limits_{a=1}^n\,\bar{U}_a U_a=1$. It is reasonable to call it the `Hopf gauge', since due to the residual $U(1)$ gauge invariance it makes manifest the structure of the Hopf fibration $S^{2n-1}\to \CP^{n-1}$. Now, the Lagrangian $\mathscr{L}={1\over \kappa}\,|\bar{D} U|^2$ in the Hopf gauge is almost the canonical form of the $\CP^{n-1}$-model, but not quite. The canonical definition of the model features two terms:
\bea
\mathscr{L}_{\CP^{n-1}}={1\over 2 \kappa}\,|D_\mu U|^2={1\over 2 \kappa}\,\left(|D U|^2+|\bar{D} U|^2\right)\,,
\eea
since the metric tensor for the flat worldsheet metric $ds^2=2 dz\,d\bar{z}$ is $g=\!\scalemath{0.7}{\begin{pmatrix} 0&1\\1&0 \end{pmatrix}}$. Therefore the Lagrangian obtained from the Gross-Neveu model may be written as
\bea
\mathscr{L}=\mathscr{L}_{\CP^{n-1}}+{1\over 2\kappa}\,\left(|\bar{D} U|^2-|D U|^2\right)
\eea
It turns out that the difference is a \emph{topological} term\footnote{The fact that  the $B$-field is topological is characteristic of symmetric space models. In general it is a non-closed 2-form.}, proportional to $\sum\limits_{a=1}^n\,\epsilon_{\mu\nu} D_{\mu}\bar{U}_a D_{\nu} U_a$, i.e. to the pull-back of the Fubini-Study form $\Omega=i\,\sum\limits_{a=1}^n\,d\bar{U}_a\wedge dU_a$. Therefore classically the models are equivalent, whereas quantum mechanically one could study the effects of adding topological terms to either of the Lagrangians.

\subsection{Quantum mechanical reduction.}\label{QMsec}

As a direct application of the formalism described above, in this section we consider the mechanical reduction of the Gross-Neveu system, which amounts to setting $\dd=\bd={d\over d t}$:
\bea
\mathscr{L}_{M}=V \cdot {\bar{D}U\over dt}+\bar{U} \cdot {D \bar{V}\over dt}+\kappa\,(\bar{U} U) (\bar{V} V)\,,\quad {\bar{D}U\over dt}={d U\over dt}-i \,\bar{\mathcal{A}}\,U \,.
\eea
The model has a gauge symmetry $U\to \lambda(t)\,U, \,V\to {1\over \lambda(t)}\,V$, where $\lambda(t)$ is a complex-valued function. Variation of the Lagrangian w.r.t. the gauge fields $\mathcal{A}, \bar{\mathcal{A}}$ gives $V\cdot U=\bar{U}\cdot \bar{V}=0$. Since the canonical commutation relations are $[U_i, V_j]=\delta_{ij},\, [\bar{U}_i, \bar{V}_j]=-\delta_{ij}$, in quantum theory these constraints translate to the following conditions on the wave function $\psi(U, \bar{U})$:
\bea\label{homfunc}
\sum\limits_{i=1}^n\,U_i {\dd \psi\over \dd U_i}=\sum\limits_{i=1}^n\,\bar{U}_i {\dd \psi\over \dd \bar{U}_i}=0
\eea
The fact that $\psi$ is annihilated by the Euler vector field $\sum\limits_{i=1}^n\,U_i {\dd \over \dd U_i}$ means that it is a homogeneous function of $U_i$ and of $\bar{U}_i$, therefore it may be regarded as a function on the projective space $\CP^{n-1}$. The Hamiltonian is
\bea\label{ham1}
H=\left(\sum\limits_{i=1}^n\,U_i \bar{U}_i\right)\,\sum\limits_{j=1}^n\,{\dd^2 \over \dd U_j \dd \bar{U}_j}
\eea
Notice that the problem of ordering of the $U$ and $V$ operators does not arise, as the possible ambiguity is proportional to a combination of operators $\sum\limits_{i=1}^n\,U_i {\dd \over \dd U_i}$ and $\sum\limits_{i=1}^n\,\bar{U}_i {\dd \over \dd \bar{U}_i}$, which annihilate the wave function anyway.

We will now diagonalize the Hamiltonian~(\ref{ham1}). To this end we will look for eigenfunctions of the form
\bea
|M\rangle:={1\over |U|^{2M}}\sum\,\psi_{i_1\ldots i_M|j_1\ldots j_M}\,U_{i_1}\cdots U_{i_M} \bar{U}_{j_1}\cdots \bar{U}_{j_M}
\eea
We will impose an additional constraint that the wave function $\psi$ is traceless w.r.t. any pair of (holomorphic/anti-holomorphic) indices: $\psi_{k i_2\ldots i_M|k j_2\ldots j_M}=0$ (in this case the operator $H$ yields zero when acting on the numerator alone). It is easy to check that the state $|M\rangle$ so defined is an eigenstate for an arbitrary wave function $\psi$:
\bear\label{Hspectrum0}
&&H\,|M\rangle=\Lambda_M\,|M\rangle,\quad\quad\textrm{where}\\ \label{Hspectrum}&& \Lambda_M=M(M+1)-M n-2M^2=-M(n+M-1)
\eear
Clearly, the huge degeneracy is due to the $SU(n)$ symmetry of the problem.

The Hamiltonian $H$, which  is the Laplace-Beltrami operator on $\CP^{n-1}$, can be brought to the somewhat more familiar form if one passes to the inhomogeneous coordinates. For example, in case of a sphere ($n=2$), setting $Z:=\frac{U_1}{U_2}$ one obtains the canonical form of the Laplacian $H=(1+|Z|^2)^2\frac{\dd^2}{\dd Z \dd \bar{Z}}$. The formula~(\ref{Hspectrum}) for the eigenvalues reduces to the familiar spectrum $\Lambda_M=-M(M+1)$.

\section{Complex symplectic phase spaces and quiver varieties}\label{compsympsec}

This completes the derivation of the equivalence between the sigma model and the Gross-Neveu model in the $\CP^{n-1}$-case. One striking consequence of this equivalence is that the sigma model turns out to be a theory with polynomial interactions. This is in sharp contrast with the conventional approach that involves nonlinear constraints such as $\bar{U}\cdot U=1$, which  subsequently lead to infinite series of interactions for the `pion' fields. The equivalence can be extended further to incorporate Grassmannian and flag manifold models, as well as an even more general class of models with complex symplectic \emph{phase} spaces. The fact that complex symplectic geometry is relevant here can already be seen at the level of the Lagrangian~(\ref{lagr2}), where the phase space is described by the kinetic term. If one forgets the gauge field for the moment, the term 
\bea\label{lagr4}
\mathscr{L}_0=\sum\limits_{a=1}^n\,V_a \bar{D} U_a\,,
\eea
which defines the so-called `$\beta\gamma$-system'~\cite{Nekrasov}, should be seen as the pull-back of the Poincar\'e-Liouville one-form $\theta_0$ corresponding to the complex symplectic form $\Omega_0:=\sum\limits_{a=1}^n \,dV_a\wedge dU_a$ on $\CC^{2n}=T^\ast \CC^n$ (by definition, $d\theta_0=\Omega_0$). The effect of the gauge field is in implementing the complex symplectic reduction by $\CC^\ast$. The resulting phase space is
\bea
\Phi=T^\ast \CC^n\!\sslash\!\CC^\ast\,.
\eea
In particular, the variation of the Lagrangian w.r.t. the gauge field gives the complex moment map constraint $\upmu_0:=\sum\limits_{a=1}^n\,V_a U_a=0$. The elementary Lagrangian~(\ref{lagr4}) is invariant w.r.t. complex transformations $U\to g\circ U, V\to V\circ g^{-1}$, where $g\in GL(n,\CC)$. Since the Lagrangian is the pull-back of a Poincar\'e-Liouville one-form, the corresponding Noether current is really the moment map for the group action:
\bea\label{muUV}
\upmu=U\otimes V\in \mathfrak{gl}_n^\ast \simeq \mathfrak{gl}_n\,.
\eea
Due to the constraint $\upmu_0=0$ of symplectic reduction, the image of the moment map is contained in the nilpotent variety $\upmu^2=0$ in $\mathfrak{sl}_n$ (since $\mathrm{Tr}(\upmu)=\upmu_0=0$). In our case this variety is the closure of the minimal nilpotent orbit $\bar{\mathcal{O}}$. Another way of seeing the relation between the nilpotent orbit and the projective space on a specific example is to parametrize $\upmu\in\mathfrak{sl}_2$ as $\upmu=\scalemath{0.7}{\begin{pmatrix} x&y\\z&-x \end{pmatrix}}$, so that $\upmu^2=0$ gives $x^2+yz=0$, which is the equation of an $A_1$ surface singularity. It may be resolved by blowing up a sphere at the origin, and the resolved space is $T^\ast \CP^1$, whereas in the $\mathfrak{sl}_n$ case it is known that the analogous resolution is $T^\ast \CP^{n-1}$~\cite{Fu}.

As a corollary of this discussion, we may rewrite the system~(\ref{lagr2}) in a more conceptual form:
\bea\label{lagr5}
\mathscr{L}=V\cdot \bar{D} U+ \bar{U}\cdot D \bar{V} +\kappa\, \mathrm{Tr}(\upmu\bar{\upmu})\,.
\eea
In the absence of the interaction term, when $\kappa=0$, the system is invariant w.r.t. the complex symmetry group $GL(n, \CC)$, and the classical solutions are the holomorphic curves in the complex phase space $\Phi$. Clearly, the coupling between $\upmu$ and $\bar{\upmu}$ breaks the symmetry down to the unitary subgroup $U(n)$, and the classical solutions are then harmonic maps into $\CP^{n-1}$, as one expects from a sigma model.

Given the Lagrangian~(\ref{lagr5}), it is now easy to see what the general class of models should be. First of all, one should replace $\Phi$ by a general complex symplectic quiver variety. In fact, the Lagrangian~(\ref{lagr5}) makes sense for an arbitrary complex symplectic variety with an action of a global symmetry group, however the system will only have a nice algebraic structure (such as polynomial interactions) in the case of quiver varieties, so it is natural to restrict to those.  In particular, the fields $U$ and $V$ in~(\ref{lagr5}), which physically are the `Weyl spinors' of the Gross-Neveu model, correspond to the $\mathrm{Hom}(L_1, L_2)$ and $\mathrm{Hom}(L_2, L_1)$ arrows of the quiver, which may be naturally paired as in the kinetic term in~(\ref{lagr5}):
\begin{center}
\begin{tikzpicture}[
baseline=-\the\dimexpr\fontdimen22\textfont2\relax,scale=1]
\draw [-stealth, OliveGreen!50, line width=2pt] (2.1,-0.12) arc [radius=1.6, start angle=60, end angle= 45];
\draw [OliveGreen!50, line width=2pt] (2.1,-0.12) arc [radius=1.6, start angle=60, end angle= 40];
\draw [-stealth, OliveGreen!50, line width=2pt] (2.52,-0.5) arc [radius=1.5, start angle=-40, end angle= -55];
\draw [OliveGreen!50, line width=2pt] (2.52,-0.5) arc [radius=1.5, start angle=-40, end angle= -60];

\draw [-stealth, OliveGreen!50, line width=2pt] (4.05,-0.45) arc [radius=1.5, start angle=140, end angle= 125];
\draw [OliveGreen!50, line width=2pt] (4.05,-0.45) arc [radius=1.5, start angle=140, end angle= 120];
\draw [-stealth, OliveGreen!50, line width=2pt] (4.47,-0.83) arc [radius=1.6, start angle=-120, end angle= -135];
\draw [OliveGreen!50, line width=2pt] (4.47,-0.83) arc [radius=1.6, start angle=-120, end angle= -140];

\draw [-stealth, OliveGreen, line width=2pt] (2.52,-0.5) arc [radius=1, start angle=140, end angle= 85];
\draw [OliveGreen, line width=2pt] (2.52,-0.5) arc [radius=1, start angle=140, end angle= 40];
\draw [-stealth, OliveGreen, line width=2pt] (4.05,-0.45) arc [radius=1, start angle=-40, end angle= -95];
\draw [OliveGreen, line width=2pt] (4.05,-0.45) arc [radius=1, start angle=-40, end angle= -140];

\draw[fill=blue] (2.52,-0.5) circle (3pt);
\draw[fill=blue] (4.05,-0.5) circle (3pt);

\node at (3.3,0.25) {$U$};
\node at (3.3,-1.2) {$V$};
\node at (2.52,-0.9) {\footnotesize $L_1$};
\node at (4.05,-0.9) {\footnotesize $L_2$};
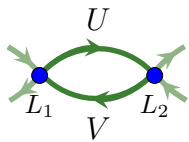
\node at (10,-0.5) {The $U$ and $V$ `Weyl spinors' as arrows of the quiver.};
\end{tikzpicture}
\end{center}

Gauge fields arise from the gauge nodes of the quiver, and $\upmu$ should be replaced by the complex moment map for (a subgroup of) the global symmetry group acting in a global node.

By a slight modification of the system~(\ref{lagr5}) one can as well obtain models with non-compact target spaces. Indeed, just as the interaction term $\mathrm{Tr}(\upmu \bar{\upmu})$ breaks the complex symmetry $GL(n, \CC)$ down to a real subgroup $U(n)$, a term $\mathrm{Tr}(\Lambda\upmu \Lambda \bar{\upmu})$ with $\Lambda=\mathrm{Diag}(-1, 1, \cdots , 1)$ a Minkowski metric would break the symmetry down to $U(1, n-1)$. The target space of the resulting model would then be $U(1, n-1)\over U(1)\times U(n-1)$ -- the complex hyperbolic space, endowed with a Bergman metric. Models of this type arise in dimensional reductions of four-dimensional gravity/matter systems, the case $n=2$ corresponding to pure gravity~\cite{BreitenlohnerMaison, GibbonsMaison, Zagermann}.

A curious property of the Lagrangian~(\ref{lagr5}) is that it seems to make no reference to a metric on the target space, which generally is one of the key ingredients of a sigma model. The metric appears only upon integration over the $V$-variables. In the case when the target space is symmetric (such as $\CP^{n-1}$ or a Grassmannian)
the metric is unique and hence coincides with the Fubini-Study metric. In other situations, for example for flag manifolds, there are families of invariant metrics, and since the Lagrangian~(\ref{lagr5}) has no free parameters, one should arrive at a special metric as a result of this procedure. Indeed, the metric that arises by this construction is the so-called normal, or reductive metric~\cite{FlagEinsteinMetrics}, which is not K\"ahler for general flag manifolds. One way to get to terms with this fact, which might seem surprising at first, is to recall that K\"ahler metrics are related to real symplectic forms, whereas the model~(\ref{lagr5}) is naturally defined in terms of complex symplectic forms. One can, in fact, define a class of models related to real symplectic forms, by replacing the $\beta\gamma$-systems -- two-dimensional field theories defined via pull-backs of complex Poincar\'e-Liouville one-forms -- with similar theories related to real one-forms (this is particularly natural for Minkowski worldsheet signature). As a result, one obtains models of Zakharov-Mikhaylov/Faddeev-Reshetikhin type~\cite{ZM2, FR}, see the recent work~\cite{Hollowood, Caudrelier, Fukushima}.

\section{Trigonometric and elliptic deformations}\label{defsec}

The $\CP^{n-1}$ sigma model, just as any other sigma model with a symmetric target space, is known to be classically integrable. This means, in particular, that the model has a Lax operator $L(z, \sigma)\in \mathfrak{g}$ depending on a spectral parameter $z$ (here $\sigma$ denotes the `spatial' coordinate on the worldsheet), which satisfies a Poisson algebra of the type~\cite{FT} (technically this is called the `ultralocal' case)
\bea\label{LPoiss}
\{L_1(z, \sigma_1)\tens{,} L_2(w, \sigma_2)\}=[r(z-w), L_1(z, \sigma_1)\otimes \mathds{1}+\mathds{1}\otimes L_2(w, \sigma_2)]\cdot \delta(\sigma_1-\sigma_2)\,,
\eea

\vspace{-0.2cm}\noindent
where $r(z-w)\in \mathfrak{g}\otimes \mathfrak{g}$ is the classical $r$-matrix satisfying the classical Yang-Baxter equation. It is well-known~\cite{BD} that such $r$-matrices are of three types: rational, trigonometric and elliptic, the spectral parameter taking values in $\CC, \CC^\ast$ or $E_\tau$ (the elliptic curve) accordingly. The homogeneous sigma model~(\ref{lagr3}) corresponds to the simplest, rational, case. The trigonometric and elliptic deformations may be easily constructed starting from the Gross-Neveu model~(\ref{lagr1}) as well\footnote{Integrability-preserving deformations of sigma models have a long history, cf.~\cite{Cherednik, Klimcik, Klimcik2, Sfetsos, DMVq}.}. All one needs to do is to insert the $r$-matrix in the quartic interaction, as follows:
\bea\label{lagrdef}
\mathscr{L}=\bar{\Psi_a} \slashed{D} \Psi_a + r(\lambda)^{cd}_{ab}\,\left(\bar{\Psi_a}{1+\gamma_5\over 2}\Psi_c\right) \,\left(\bar{\Psi_d}{1-\gamma_5\over 2}\Psi_b\right)\,,
\eea
In contrast to~(\ref{LPoiss}), here we have written the $r$-matrix as an element of $\mathfrak{g}\otimes \mathfrak{g}^\ast$, using the Killing metric to raise or lower the indices (we use the two notations interchangeably, depending on the context). The theory now depends on a parameter $\lambda$, which should be understood as the deformation parameter. For brevity, in writing down explicit expressions, we restrict to the case of $\mathfrak{sl}_2$~\cite{FT}:
\bear\label{rtrig}
&&\hspace{-2cm}\mathrm{Trigonometric}:\quad\quad  r(\lambda)={1\over 2 \sinh{\lambda}}\left(\sigma_1\otimes \sigma_1+\sigma_2\otimes \sigma_2+\cosh{\lambda} \;\sigma_3\otimes \sigma_3\right)\\ \label{rell}
&&\hspace{-2cm}\mathrm{Elliptic}:\quad\quad\quad  r(\lambda)={i\over 2\, \mathrm{sn}\,(i \lambda)}\left(\sigma_1\otimes \sigma_1+ \mathrm{dn}\, (i \lambda)\;\sigma_2\otimes \sigma_2+\mathrm{cn}\,{(i \lambda)} \;\sigma_3\otimes \sigma_3\right)
\eear
The rational $r$-matrix, which enters the undeformed model, is recovered in the limit $\lambda\to 0$. The Jacobi elliptic functions $\mathrm{sn}, \mathrm{cn}, \mathrm{dn}$ depend on an additional variable -- the elliptic modulus, for example $\mathrm{sn}(i \lambda)=\mathrm{sn}(i \lambda, k)$, and the trigonometric functions are restored in the limit $k\to 0$.

\begin{figure}
\centering
\bea\nonumber
\begin{tikzpicture}[
baseline=-\the\dimexpr\fontdimen22\textfont2\relax,scale=1.2]
\draw [-stealth, red!50, line width=2pt] (2,0) arc [radius=1, start angle=100, end angle= 120];
\draw [-stealth, red!50, line width=2pt] (2,0) arc [radius=1, start angle=100, end angle= 180];
\draw [-stealth, red!50, line width=2pt] (2,0) arc [radius=1, start angle=100, end angle= 255];
\draw [red!50, line width=2pt] (2,0) arc [radius=1, start angle=100, end angle= 260];
\draw [-stealth, blue!50, line width=2pt] (1,0) arc [radius=1, start angle=80, end angle= 60];
\draw [-stealth, blue!50, line width=2pt] (1,0) arc [radius=1, start angle=80, end angle= 0];
\draw [-stealth, blue!50, line width=2pt] (1,0) arc [radius=1, start angle=80, end angle= -75];
\draw [blue!50, line width=2pt] (1,0) arc [radius=1, start angle=80, end angle= -80];
\node at (2.2,0) {\footnotesize $k$};
\node at (0.8,0) {\footnotesize $i$};
\node at (0.8,-2) {\footnotesize $j$};
\node at (2.2,-2) {\footnotesize $l$};
\draw [-stealth, red!50, line width=2pt] (5,-1.95) arc [radius=1, start angle=260, end angle= 240];
\draw [-stealth, red!50, line width=2pt] (5,-1.95) arc [radius=1, start angle=260, end angle= 180];
\draw [-stealth, red!50, line width=2pt] (5,-1.95) arc [radius=1, start angle=260, end angle= 110];
\draw [red!50, line width=2pt] (5,-1.95) arc [radius=1, start angle=260, end angle= 100];
\draw [-stealth, blue!50, line width=2pt] (4,0) arc [radius=1, start angle=80, end angle= 60];
\draw [-stealth, blue!50, line width=2pt] (4,0) arc [radius=1, start angle=80, end angle= 0];
\draw [-stealth, blue!50, line width=2pt] (4,0) arc [radius=1, start angle=80, end angle= -75];
\draw [blue!50, line width=2pt] (4,0) arc [radius=1, start angle=80, end angle= -80];
\node at (5.2,0) {\footnotesize $l$};
\node at (3.8,0) {\footnotesize $i$};
\node at (3.8,-2) {\footnotesize $j$};
\node at (5.2,-2) {\footnotesize $k$};
\end{tikzpicture}
\eea
\caption{Diagrams contributing to the $\beta$-function at one loop. The red and blue arrows are the $\langle U\, V\rangle$ and $\langle \bar{U}\, \bar{V}\rangle$ Green's functions. The quartic vertex is given by the classical $r$-matrix $r(\lambda)^{cd}_{ab}$.} \label{figbeta}
\end{figure}
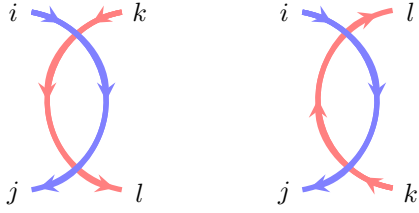

A natural question to ask is whether the deformed model is renormalizable. Indeed, out of a potentially huge number of couplings (in the $\mathfrak{sl}_n$-case of order $O(n^4)$), the $r$-matrix provides a one-parametric subfamily depending on $\lambda$, so one would like to prove that this subfamily is stable under renormalization. The one-loop $\beta$-function is given by two simple diagrams shown in Fig.~\ref{figbeta} (note that no background field method is necessary in this approach):
\bea
\beta_{ij}^{kl}(\lambda)=\sum\limits_{p, q=1}^{n}\,\left(r(\lambda)_{ip}^{kq} r(\lambda)_{pj}^{ql}-r(\lambda)_{ip}^{ql} r(\lambda)_{pj}^{kq}\right)\,.
\eea
The RG-flow equation is ${d\over d\tau}\,r_{ij}^{kl}=\beta_{ij}^{kl}$, and in the $n=2$ case~(\ref{rtrig})-(\ref{rell}) for both types of $r$-matrices the solution is $\lambda =2 \tau$. The trigonometric case was studied\footnote{The $r$-matrix used in that paper is related to~(\ref{rtrig}) by a similarity transformation. As a consequence of this, in~\cite{BykovGN} an additional coordinate transformation was needed in order to satisfy the RG-equation, which is not necessary in the present formulation.} for general~$n$ in~\cite{BykovGN}, and the solution was found to be
\bea
\lambda=\lambda_0+n\,\tau\,,
\eea
i.e. the flow of the additive `spectral variable' $\lambda$ is linear. In the rational case, when $\lambda$ is related to the radius of target space $R$ as $\lambda=R^2$, the same solution also holds, as it coincides with the standard $\beta$-function of the homogeneous sigma model (cf.~\cite{ZaremboLect} or the books~\cite{ZJBig, KetovBook}). Therefore it is natural to expect that this is a universal solution for all three cases.

The trigonometric case is especially illuminating. Here the RG-time variable $\tau$ is defined on a semi-axis, so that $\lambda\in(0, \infty)$, and one has a so-called `ancient' solution of the Ricci flow. The limits $\lambda \to \infty$ and $\lambda \to 0$ correspond to the UV- and IR- limits respectively. In the IR-limit the metric becomes round (i.e. Fubini-Study), with a vanishing radius, which is a sign of an infrared singularity. In the UV the metric becomes flat, and the space degenerates to an $(n-1)$-dimensional cylinder $(\CC^\ast)^{n-1}$. The deformed Gross-Neveu model~(\ref{lagrdef}) should be thought of as a multi-dimensional generalization of the `sausage' model~\cite{Onofri}. The latter arises in the $n=2$ case with the trigonometric $r$-matrix~(\ref{rtrig}), and the corresponding RG-flow interpolates between the round 2-sphere and a cylinder.

The remarkable simplicity of the solutions is likely a consequence of the integrability of the deformed model\footnote{The relation between integrability and renormalizability has been discussed throughout the years, cf.~\cite{Fateev, Onofri, ValentKlimcik, Lukyanov, Hoare1, Hoare2, DelducRG}.}~(\ref{lagr4}). As mentioned above, one of the hallmarks of integrability is the existence of a `Lax pair', or equivalently of a one-parametric family of connections~$A(\nu)$, whose flatness encodes the e.o.m. of the model. For the present models the connection is
\bea\label{Auconn}
A(\nu)=\left(r(\nu)\circ\upmu\right)\;dz-\left(r(\nu-\lambda)\circ\bar{\upmu}\right)\;d\bar{z}\,,
\eea
where the circle $\circ$ indicates the contraction of $r\in \mathfrak{g}\otimes \mathfrak{g}^\ast$ with an element of $\mathfrak{g}$. The spectral parameter~$\nu$, as well as the deformation parameter $\lambda$, take values in a complex curve, which is $\CC, \CC /\mathds{Z}$ or $\CC/ \mathds{Z}^2$ depending whether one has a rational, trigonometric or elliptic case. This form of the connection may be obtained through the derivation of~\cite{CYa}, starting from a four-dimensional Chern-Simons theory coupled to surface defects.

As one can find from the explicit expression~(\ref{muUV}) and from the Lagrangian~(\ref{lagrdef}), the moment maps satisfy the following equations of motion:
\bea\label{mueom1}
\bd\upmu=\, [\upmu, r(-\lambda)\circ \bar{\upmu}]\,,\quad\quad \dd \bar{\upmu}=\,[\bar{\upmu},\; r(\lambda)\circ\upmu]\,.
\eea
In the undeformed case these are the familiar e.o.m. of the principal chiral model, which suggests that models of our class are certain reductions of it (this line of thought is pursued in more detail in~\cite{BykovNilp}). As it follows from~(\ref{Auconn}) and~(\ref{mueom1}), the flatness condition for $A(\nu)$ is quadratic in the $r$-matrices, and it turns out that this quadratic relation is the classical Yang-Baxter equation. Ultimately this is the reason that we have chosen a classical $r$-matrix for the quartic coupling in the Lagrangian~(\ref{lagrdef}).

It is worth mentioning, that even in the homogeneous limit $\lambda, \nu\to 0$, the connection $A(\nu)$ is different from the canonical Pohlmeyer connection~\cite{Pohlmeyer} $\tilde{A}(\tilde{\nu})$ for the $\CP^{n-1}$ sigma model. The two connections are related by a local gauge transformation $\tilde{A}(\tilde{\nu})=GA(\nu) G^{-1}-Gd G^{-1}$ and a redefinition of the spectral parameter $\tilde{\nu}=\tilde{\nu}(\nu)$~\cite{BykovZeroCurv}, however only the connection $A(\nu)$ leads to ultralocal Poisson brackets of the form~(\ref{LPoiss}). For the case of symmetric space models (such as $\CP^{n-1}$) this was first shown in~\cite{Bytsko, Zagermann}; an analogous situation has been shown to hold in the case of $\mathbb{Z}_m$-symmetric target spaces~\cite{DelducZT}; in the deformed case the ultralocal Lax pair for $\CP^1$ (the `sausage') has been derived in~\cite{Kotousov}.

\section{Models with fermions and the supersymplectic quotient}\label{fermionsec}

The next question is how to include fermions in our models. After all, it is the fermionic version of the Gross-Neveu model that was originally studied in the~70's. It turns out that all known integrable fermionic extensions of the $\CP^{n-1}$ model may be naturally described in the formalism described above. To treat these cases, first one incorporates fermions in the phase space by replacing
\bea
\Phi_0=T^\ast \CC^n \quad\quad \longmapsto \quad\quad \widetilde{\Phi_0}=T^\ast \CC^{n|n}\,.
\eea
If we only had the free Lagrangian~(\ref{lagr4}), this replacement would lead to a global symmetry group $GL(n|n)$ in place of $GL(n)$. Just as in the bosonic case, the interaction term, defined through the moment map $\upmu$ of a global symmetry subgroup $G_{\mathrm{gl}}\subset GL(n|n)$, reduces the symmetry. What exactly the remaining symmetry group is depends on what~$G_{\mathrm{gl}}$ we choose. For example, similarly to the bosonic case, if we keep the full group $G_{\mathrm{gl}}= GL(n|n)$, the interaction term reduces the symmetry to its unitary subgroup $U(n|n)$. This means that the model has target-space supersymmetry, and in this case the target space is the super-projective space $\CP^{n-1|n}$. For most applications, however, one would like to evade target-space supersymmetry: for example, in the worldsheet supersymmetric $\CP^{n-1}$ model we expect the symmetry group to be the same as in the bosonic model, i.e. $U(n)$ (up to an additional $R$-symmetry group).

One way to reduce the symmetry would be to choose a smaller group $G_{\mathrm{gl}}$, such as $G_{\mathrm{gl}}=SL(n, \CC)$, where $SL(n, \CC)$ is the subgroup of $GL(n|n)$ that rotates the bosonic variables. This leads to the $\CP^{n-1}$ model with minimally coupled fermions. Another possibility is to choose $G_{\mathrm{gl}}=SL(n, \CC)$ diagonally embedded in $GL(n|n)$, which turns out to be the relevant setup to obtain the model with worldsheet supersymmetry. The corresponding moment map has the form
\bea\label{muBC}
\upmu:=U\otimes V-C\otimes B\,,
\eea
where $C$ and $B$ are the fermionic coordinates in $\widetilde{\Phi_0}$. Incidentally, (\ref{muBC}) is invariant w.r.t. $GL(1|1)$-transformations $\mathrsfso{U}\to g \circ \mathrsfso{U}$, $\mathrsfso{V}\to \mathrsfso{V}\circ g^{-1}$ that rotate the doublets
\bea
   \mathrsfso{U}:=\begin{pmatrix} 
      U  \\
      C  \\
   \end{pmatrix},\quad\quad \mathrsfso{V}:=\begin{pmatrix}
      V  &
      B  
   \end{pmatrix}\,.
\eea
This means that part of the target-space supersymmetry still remains in the interacting theory. The way to eliminate it is to gauge part of it. As shown in~\cite{BykovSUSY}, one should gauge a subgroup $G_{\triangle}\subset GL(1|1)$ comprising matrices of the form
\bea\label{trianglesub}
G_{\triangle}:=\left\{\quad g\in SL(1|1)\,:\quad g=\begin{pmatrix} 
      \uplambda & 0  \\
      \upchi &  \uplambda
   \end{pmatrix}\quad \right\}
\eea
Notice that $\uplambda$ is a bosonic variable living in $\CC^\ast$, and $\upchi$ parametrizes a fermionic subgroup. As a result, the phase space is a super-variety obtained by the super-symplectic reduction:
\bea\label{SUSYred}
\Phi=T^\ast \CC^{n|n}\!\sslash\! G_{\triangle}\,.
\eea 
One way to understand the meaning of the triangular subgroup $G_{\triangle}$ is to consider the way it acts on the `configuration space' $\CC^{n|n}$. The action  is $\mathrsfso{U}\to g\cdot \mathrsfso{U}$, which in components is $U\to \uplambda\, U, C\to \uplambda\, C+\upxi \,U$. Clearly, the $U$-variables describe the projective space $\CP^{n-1}$, and the $C$-variables take values in a vector bundle over that projective space. Taking the quotient by multiples of $U$ means one has a quotient $\CC^n / \mathcal{O}(-1)$ of the trivial bundle. An additional multiplication by $\uplambda$ means that one has the bundle $V=\mathcal{O}(1)\otimes (\CC^n / \mathcal{O}(-1))$, which is the tangent bundle $V=T \CP^{n-1}$. This is in line with general principles, since fermions in supersymmetric theories take values in the tangent bundle.

In comparing the above construction to the conventional formulation of the SUSY sigma model, the following observation is illuminating.~The standard formulation~\cite{AddaSUSY} involves an $n$-tuple of Dirac fermions $\Theta_a$ ($a=1, \ldots, n$), with an additional constraint of orthogonality to the vector of bosonic variables:
\bea\label{FermBosOrtho}
\bar{\Theta}\circ U=0\,.
\eea
In our formulation, we decompose the Dirac fermion into Weyl spinors in exactly the same way as we did for the bosonic variables~(\ref{psidecomp}): $\Theta=\begin{pmatrix} C\\ \bar{B}\end{pmatrix}$. As a result, the constraint~(\ref{FermBosOrtho}) disassembles into two: $B\circ U=0$ and $\bar{C}\circ U=0$. In the approach based on the supersymplectic quotient~(\ref{SUSYred}), the first of these constraints is the holomorphic \emph{moment map} for the action of the one-parametric fermionic subgroup of $G_{\triangle}$ (parametrized by $\upchi$ in~(\ref{trianglesub})). The second one, in turn, should be viewed as a \emph{gauge choice} for taking the quotient w.r.t. this one-parametric subgroup. To summarize, the constraint~(\ref{FermBosOrtho}) is a result of two steps of the symplectic reduction by a fermionic~group.

One should also point out that in the supersymmetric case the Gross-Neveu formulation is not totally unexpected, as it has long been known that `supersymmetrization' involves coupling the bosonic sector to a chiral Gross-Neveu model. What is remarkable, however, is that the bosonic part is a chiral Gross-Neveu model itself, albeit a bosonic one!

\section{Outlook}

In this mini-review we discussed the salient features of the correspondence between sigma models of a special class (which includes models with complex homogeneous target spaces) and chiral Gross-Neveu models. We believe that this correspondence is fundamental for the proper understanding of such theories, both in the classical and quantum domains. Many questions remain open, among them the analysis of potential higher-loop corrections to the $\beta$-function, studying ways of coupling Gross-Neveu models to gravity, reaching a better understanding of the relation to Ashtekar variables~\cite{Zagermann, BykovAffleck} and dimensional reductions of four-dimensional gravity, and ultimately constructing  solutions of the quantum models (the pure fermionic case was solved in~\cite{Andrei, Destri}). 

\vspace{1cm}
\noindent
\textbf{Acknowledgments.} I would like to thank A.~A.~Slavnov for support, as well as A.~K.~Pogrebkov for the invitation to give a talk at the ``Polivanov-90'' conference. Some of this material was also presented at the conferences `GLSMs-2020' (Virginia Tech, USA) and `RAQIS-2020' (Annecy, France), and I am grateful to the organizers E.~Sharpe and E.~Ragoucy, respectively, for the invitations. This work has been supported by Russian Science Foundation grant RSCF-20-72-10144.

\makeatletter
\renewcommand\@biblabel[1]{#1.}
\makeatother

{
\setstretch{0.8}
\setlength\bibitemsep{3pt}
\printbibliography
}

\end{document}